%% file: template.tex
\documentclass[cameraready]{Interspeech}

\usepackage{multirow}
\usepackage{xcolor}
\usepackage{amssymb}
\usepackage{amsfonts}
\usepackage{diagbox}
\usepackage{booktabs}
\usepackage{xfrac}
\usepackage{tabularx}
\usepackage{amsmath}
\usepackage{amssymb}
\usepackage{tikz}
\usetikzlibrary{fit}
\usepackage{url}
\usepackage{hyperref}

\usepackage[table]{xcolor}
\usepackage{tabularx}
\newcolumntype{Y}{>{\centering\arraybackslash}X}
\usepackage{makecell}

\input{preamble}

\title{\ours: Text-Guided Instrument Timbre Transfer with Target-Adaptive Structural Control}

\author[affiliation={1}, orcid=0009-0005-6847-2623]
{Dabin}{Kim}
\author[affiliation={2}, orcid=0000-0002-4589-6061]{Junwon}{Lee}
\author[affiliation={1,2}, orcid=0000-0003-2664-2119]{Juhan}{Nam}

\address{
    \textsuperscript{1}Graduate School of Cultural Technology and \textsuperscript{2}Graduate School of AI, KAIST, South Korea
}

\email{dabinchi9598@kaist.ac.kr, james39@kaist.ac.kr, juhan.nam@kaist.ac.kr}


\usepackage{comment}

\begin{document}

\maketitle

\input{sec/0_abstract}

\input{sec/1_introduction}
\input{sec/2_method}
\input{sec/3_experiment}
\input{sec/4_conclusion}

\section{Generative AI Use Disclosure}
Generative AI tools were used only for camera-ready compliance checking, formatting, and minor editorial polishing. They were not used to generate scientific content, experimental results, or conclusions. The authors have reviewed the manuscript and take full responsibility for its content.

\bibliographystyle{IEEEtran}
\bibliography{mybib}

\end{document}

%% file: preamble.tex
\definecolor{softBlue}{RGB}{190, 220, 255}
\definecolor{softGreen}{RGB}{190, 240, 200}
\definecolor{softRed}{RGB}{255, 210, 210}
\newcommand{\cB}[1]{\cellcolor{softBlue!#1}}

\newcommand{\cR}[1]{\cellcolor{softRed!#1}}

\newcommand{\ours}{AdaTT\xspace}

%% file: sec/0_abstract.tex
\begin{abstract}
    This paper addresses timbral ambiguity 
    in instrument timbre transfer under fine-grained structural conditions. We argue this issue
    stems from instrument-specific expressive details in these conditions, which conflict with the target timbral properties.
    For example, imposing a violin's pitch-dominant vibrato contours onto a flute, which naturally exhibits loudness-dominant vibrato, impairs
    timbral fidelity.
    We propose \textbf{\ours}, a target-adaptive system that ensures high timbral fidelity across diverse timbre transfer scenarios
    within the ControlNet scheme. It selectively scales the frame-wise influence of pitch and loudness controls via text prompts to match the target instrument's identity. We also present 
    a semi-automatic
    data construction pipeline 
    to teach the model which expressive details to transform or preserve. Results show \ours achieves superior timbral fidelity and naturalness while retaining score-level content. Audio samples are available at \url{https://dabinkim0.github.io/adatt/}.
    \end{abstract}

%% file: sec/1_introduction.tex
\section{Introduction}

Timbre transfer is a task that transforms the instrumental identity of audio tracks while preserving core musical content such as melody and rhythm.
This capability provides a versatile framework for composition and arrangement,
enabling non-proficient users to overcome practical limitations---ranging from a lack of instrumental expertise to financial constraints---and realize their creative intent.

A primary challenge in this task arises from the inherently ambiguous nature of timbre, which complicates the distinction between acoustic attributes to be replaced and those to be retained.
Psychoacoustic studies~\cite{Grey77-Multidimensional, McAdams95-Perceptual} underscore this complexity, indicating that timbre encompasses not only time-invariant harmonic patterns but also time-varying expressive details, such as attack transients and vibrato, deeply rooted in instrument-specific sound production mechanisms.
For example, the gradual onset and pitch-dominant vibrato of sustained violin tones constitute distinct instrumental signatures, analogous to how prosodic contours in speech form
a key component of speaker identity.
Hence,
to clearly define the objective of high-fidelity timbre transfer,
we interpret the underlying temporal structure (e.g., pitch and loudness shapes) of performance audio as a mixture of instrument-agnostic \textit{score-level content} and instrument-specific \textit{expressive details}.
Under this formulation, natural timbre transfer requires preserving score-level content while adapting original expressive details to align with the target instrument's identity.

\begin{figure}[t!]
  \centering
  \includegraphics[width=\linewidth]{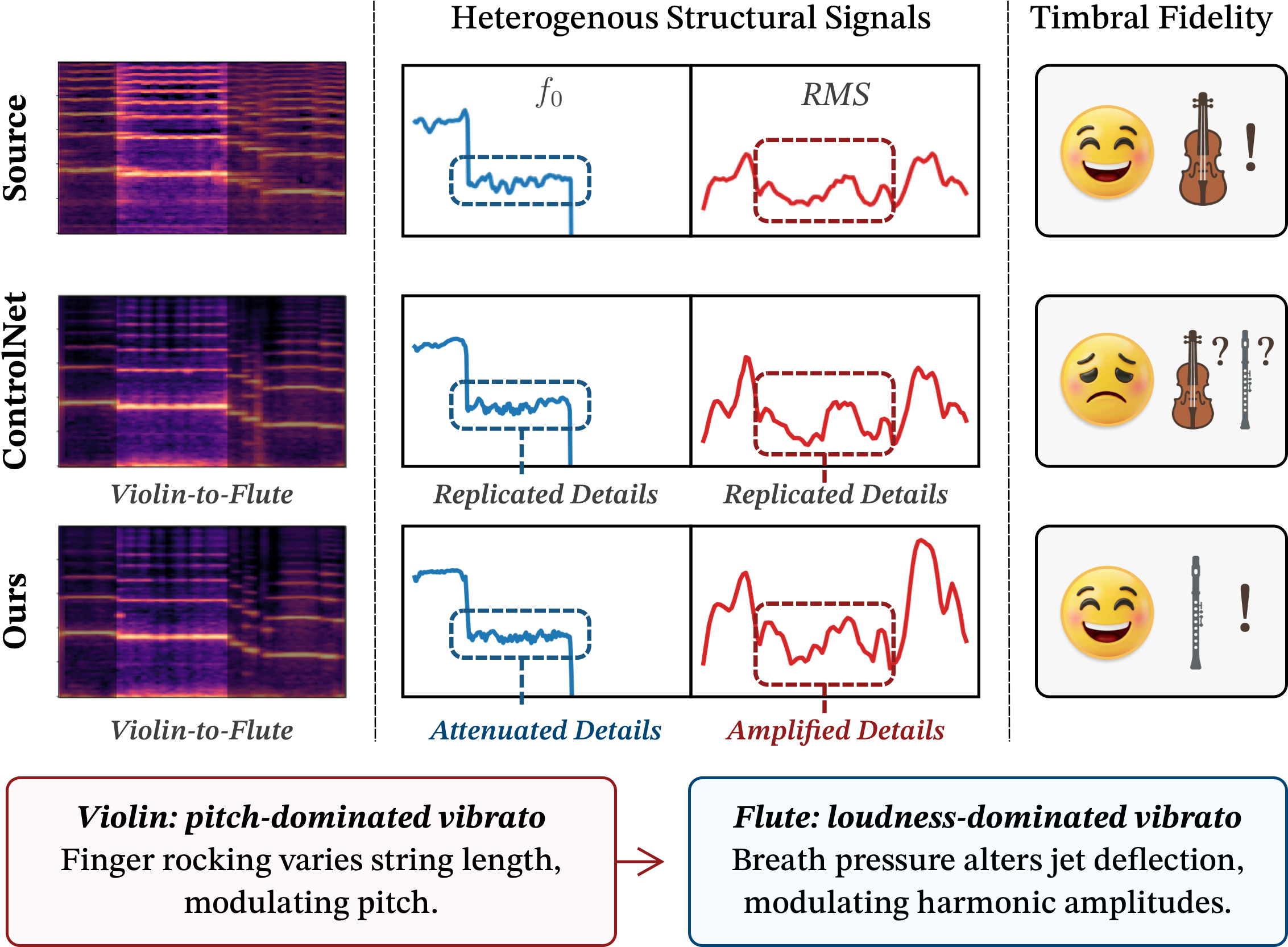}
  \caption{Visual comparison of Violin-to-Flute timbre transfer.
  }
  \label{fig:core_concept}
  \vspace{-8pt}
\end{figure}

Existing deep generative methods, including both training-from-scratch methods~\cite{Luo19-GMVAE, Nakashima22-Hyperbolic, Cifka21-SSVQVAE, Zhang25-Vevo, Engel20-DDSP, Huang18-TimbreTron} and recent paradigms that leverage text-to-music (TTM)~\cite{Liu23-AudioLDM, Liu23-AudioLDM2, Evans24-StableAudioOpen} generative priors for language-driven control and high-fidelity synthesis, have struggled to handle source expressive details that are incompatible with
the target instrument's identity.
Most of them fail to decouple these nuances from score-level content and indiscriminately preserves both irrespective of
source--target relationships, which result in timbral ambiguity and unnatural artifacts.
To closely examine this issue within TTM-based approaches,
we categorize their structural preservation strategies into two prominent paradigms:
(1) \textit{ControlNet-based finetuning}~\cite{Wu24-MusicControlNet, Hou24-SAO-ControlNet, Baker25-LiLAC}
injects source-extracted structural signals as explicit conditions via trainable adapters, guiding the frozen TTM backbone to reconstruct the original audio.
However, this reconstruction-driven objective compels the model to rigidly replicate the source's fine-grained expressive details, even in zero-shot timbre transfer cases
where the source and target instruments differ.
This naive adherence pulls the generated result away from the natural manifold of the target timbre, often forcing a reliance on coarser representations like 12-tone equal temperament (12-TET) melodic sequences~\cite{Wu24-MusicControlNet} at the expense of the original expressive richness (e.g., articulations and performative dynamics).
(2) Alternatively, \textit{inference-time editing}~\cite{Zhang24-MusicMagus, Manor24-ZETA, Baoueb25-DiffTONE} utilizes implicit structural guidance like attention or noise injection~\cite{Hertz22-P2P, Tumanyan23-PnP} to mitigate the rigidity of the former approach.
While this latent-level guidance better preserves timbral fidelity, it frequently suffers from structural deviations due to the lack of additional training with explicit conditioning.

To ensure robust timbral fidelity across diverse transfer scenarios,
we propose \textbf{\ours}, a target-adaptive mechanism that balances the relative strengths of dual heterogeneous structural controls to match the target timbre.
Guided by a text prompt that specifies the target instrument, it dynamically scales the pitch and loudness conditioning frame-wise within the ControlNet framework.
This enables the model to effectively preserve score-level content while flexibly adapting source expressive details to the target instrument's characteristics, rather than discarding them, thereby retaining the expressive richness of original performance audio.
We further present a semi-automatic data construction pipeline for task-specific finetuning, which teaches the model to discriminate between acoustic attributes to preserve and those to transfer without labor-intensive human annotation.
Experiments show that our method achieves superior timbral fidelity and naturalness while maintaining precise score-level content preservation comparable to the standard ControlNet baseline.

\begin{figure}[t!]
  \centering
  \includegraphics[width=0.9\linewidth]{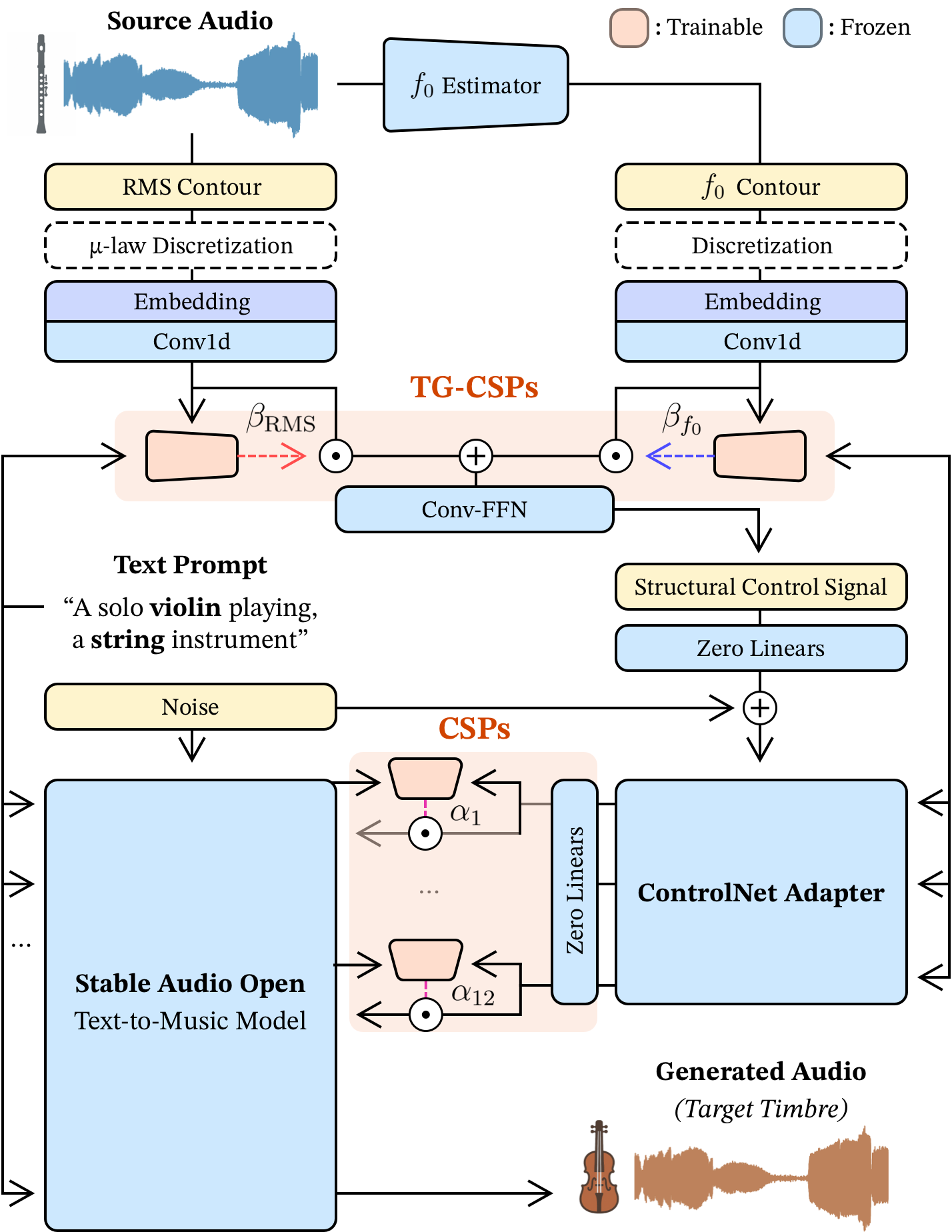}
  \caption{
  Overall architecture of \ours, integrating two sets of lightweight modules into frozen SAO-ControlNet.
  }
  \label{fig:architecture}
  \vspace{-8pt}
\end{figure}

%% file: sec/2_method.tex
\section{Method}

\subsection{Diffusion transformer with ControlNet}
We formulate timbre transfer as a conditional generation task, where a text prompt stating the target instrument serves as the conditioning input alongside the pitch and loudness shapes extracted from the source audio.
For this purpose, we employ Stable Audio Open (SAO)~\cite{Evans24-StableAudioOpen} as our generative backbone, a latent diffusion model for text-to-music generation that operates in an audio DAC-VAE~\cite{Kingma14-VAE,kumar2023highfidelity} latent space. The denoising network $v_\theta$ implemented as a Diffusion Transformer (DiT)~\cite{Peebles23-DiT} is trained via the velocity prediction~\cite{Salimans22-ProgDist} objective.

To inject source-derived structural controls into the frozen SAO, 
we use ControlNet~\cite{Zhang23-ControlNet} adapted for DiTs, following Hou~et~al.~\cite{Hou24-SAO-ControlNet}.
This framework introduces a trainable copy of the first $N$ transformer blocks, connected via zero-initialized linear layers $\mathcal{Z}$ to prevent disrupting the pretrained generative priors:
\begin{gather}
    h_{i+1} = \mathcal{F}_{\theta, i}\bigl(h_i + \mathcal{Z}(\Delta_{\phi,i}(u_i))\bigr), \\
    u_0 = h_0 + \mathcal{Z}(s)
    \label{eq:control_injection}
\end{gather}
where $h_i$ denotes the original input to the $i$-th frozen block $\mathcal{F}_{\theta, i}$, $u_i=\Delta_{\phi, i-1}(u_{i-1})$ is the input to the corresponding trainable copy $\Delta_{\phi, i}$, and $s$ denotes the structural conditioning signal.
Unlike the baseline~\cite{Hou24-SAO-ControlNet} that constructs $s$ using a single polyphonic feature (i.e., top-$k$ CQT), we use fundamental frequency ($f_0$)\footnote{$f_0$ is extracted using CREPE \cite{Kim18-CREPE}.} and Root Mean Square (RMS) contours to capture
both pitch and loudness shapes of monophonic audio.
These signals are quantized into discrete bins and embedded via learnable lookup tables, $E_{f_0}$ and $E_{\text{RMS}}$, then passed through 1D convolutional layers to yield latent control features $z_{f_0}$ and $z_{\text{RMS}}$. These features are fused via a feed-forward network (Conv-FFN) to produce the unified control signal $s$:
\begin{equation}
    s = \mathrm{ConvFFN}\bigl(z_{f_0} + z_{\text{RMS}}\bigr).
\label{eq:input_injection}
\end{equation}
We refer to this base framework as \textit{SAO-ControlNet}.

\subsection{Target-adaptive scaling for heterogeneous controls}
\label{sec:adaptive_control}
To enable adaptive modulation of 
source's expressive details during timbre transfer,
we propose \textbf{\ours}, a target-adaptive mechanism that operates within the ControlNet framework.
Unlike SAO-ControlNet, which strictly preserves fine-grained structure without considering the source--target context,
\ours dynamically scales the frame-wise local influence of each control via text prompts, transforming target-incompatible expressive details
to suit the desired instrument.
This modulation is facilitated by integrating two sets of lightweight modules: \textit{Control Scale Predictors (CSPs)} and \textit{Text-Guided CSPs (TG-CSPs)}.

The CSPs predict scale vectors $\alpha_i \in [0, 1]^{L \times 1}$ that perform frame-wise scaling on the output of the $i$-th ControlNet block $\mathcal{Z}(\Delta_{\phi,i}(u_i))$, inspired by SmartControl~\cite{Liu24-SmartControl}.
Mathematically, $\alpha_i = \mathrm{CSP}_{i}([h_i; h_i + \mathcal{Z}(\Delta_{\phi,i}(u_i))])$ where $[\cdot ; \cdot]$ denotes channel-wise concatenation and  $\mathrm{CSP}_i=\sigma(\mathcal{Z}_{\alpha}(\Phi(\cdot)))$ consists of two convolutional layers with SiLU activations $\Phi$, followed by a zero-initialized convolution $\mathcal{Z}_{\alpha}$ and a sigmoid function $\sigma$. Finally, the hidden features for the subsequent backbone layer are given by $h_{i+1} = \mathcal{F}_{\theta, i}(h_i + \alpha_i \odot \mathcal{Z}(\Delta_{\phi,i}(u_i)))$, where $\odot$ denotes element-wise multiplication.
Notably, unlike SmartControl, we initialize the bias of $\mathcal{Z}_{\alpha}$ to $+3$ (yielding $\alpha_i \approx 0.95$) to preserve the initial control strength.

As a key component of \ours, we introduce the
Text-Guided Control Scale Predictor (TG-CSP) to extend
control beyond the ControlNet output stage into its input stage.
While standard CSPs effectively regulate the overall control intensity, they may be insufficient for balancing the relative strengths of the pitch ($f_0$) and loudness (RMS) controls, which play distinct acoustic roles in shaping instrumental identity. This limitation likely arises because these heterogeneous signals become deeply entangled within the ControlNet's hidden space, hindering fine-grained, signal-specific modulation.
To circumvent this bottleneck, we apply signal-wise scaling vectors $\beta_{f_0}$ and $\beta_{\text{RMS}}$ to independently modulate each control feature $z_k$ ($k \in \{f_0, \text{RMS}\}$).
In particular, each $z_k$ is concatenated with the temporally broadcast text embedding $c_{\mathrm{txt}}$ from the T5 encoder~\cite{Raffel20-T5} to predict its corresponding frame-wise scale $\beta_k$:
\begin{gather}
    \beta_{k} = 2\cdot\sigma\Bigl(\mathcal{Z}_{\beta}\bigl(\Phi_{k}([z_{k}; c_{\mathrm{txt}}])\bigr)\Bigr),
    \label{eq:tg_csp}
    \\
    s = \mathrm{ConvFFN}\bigl(\beta_{f_0} \odot z_{f_0} + \beta_{\text{RMS}} \odot z_{\text{RMS}}\bigr),
    \label{eq:beta_fusion}
\end{gather}
where $\Phi_{k}$ denotes a stack of two 1D convolution layers with SiLU activations, and $\mathcal{Z}_{\beta}$ is a zero-initialized convolution.
Note that the zero-initialization of $\mathcal{Z}_{\beta}$ preserves the original control strength ($2\cdot\sigma(0)=1$) at the beginning of training,
while the scaled sigmoid constrains the predicted values to $[0, 2]$.
Consequently, guided by the text prompt, this formulation enables \ours
to selectively amplify or attenuate source expressive details, harmonizing heterogeneous structural controls.

\input{tab/table_data-stat}
\subsection{Cross-instrument data construction}
\label{sec:data_construction}
Training \ours for timbre transfer requires source--target audio pairs where distinct instrumental tracks share consistent score-level content yet exhibit expressive details aligned with their respective instruments.
However, obtaining such ground-truth transferred audio from real-world performances is practically infeasible, as it demands strictly controlled replication by skilled musicians.
To address this scarcity, we propose a data construction pipeline that leverages SAO-ControlNet inference, integrating metric-based filtering with expert verification to generate synthetic pairs with high perceptual naturalness.

\noindent\textbf{Pitch-range clustering.}
To minimize pitch range discrepancies, we categorized instruments into three clusters based on average $f_0$: $C_{\text{High}}$ (Violin, Flute, Oboe, Clarinet, Trumpet), $C_{\text{Mid}}$ (Viola, Bassoon, Saxophone, Horn), and $C_{\text{Low}}$ (Double Bass, Cello, Tuba, Trombone).
Using 40 samples per instrument, we formed all possible cross-instrument pairs within each cluster.

\noindent\textbf{Two-stage curated inference.}
We apply the scaling factors $\alpha$ (where $\alpha_i = \alpha, \forall i$), $\beta_{f_0}$, and $\beta_{\text{RMS}}$ to the SAO-ControlNet, performing inference for each point in the predefined parameter search grid without further training.
The optimal combination of scaling factors is identified via a per-sample, two-stage procedure to generate the final pseudo ground-truth audio, detailed as follows:
\begin{enumerate}[leftmargin=*]
    \item \textit{Inter-control balancing ($\beta$)}:
    Fixing $\alpha=1.0$, we generate samples using $(\beta_{f_0}, \beta_{\text{RMS}}) \in \{(1.8, 0.2), \dots, (0.2, 1.8)\}$. After ranking these samples by $\text{Chroma}$ scores for structural consistency, we select the top-3 candidates for expert verification to identify the optimal structure--timbre balance.
    \item \textit{Global control scaling ($\alpha$)}:
    With the optimal $(\beta_{f_0}, \beta_{\text{RMS}})$ values fixed, we then generate samples for $\alpha \in \{0.2, \dots, 1.0\}$. We filter out candidates with $\text{Chroma} < 0.7$ and select the best one to ensure maximal perceptual fidelity.
\end{enumerate}
Note that we impose the constraint $\beta_{f_0} = 2 - \beta_{\text{RMS}}$ to maintain consistent control energy relative to the default configuration of $(\beta_{f_0}, \beta_{\text{RMS}}) = (1.0, 1.0)$.
Expert verification strictly excludes candidates that fail to preserve score-level content or 
exhibit target-appropriate expressive details, including those with melody corruption, source-instrument leakage, or unnatural artifacts.
The resulting 1,321 high-quality pairs ($\approx$10\% of the SAO-ControlNet training set; 4.40 hrs) constitute the \textit{instrument transfer set}
detailed in Table~\ref{tab:dataset_stats}.

%% file: tab/table_data-stat.tex
\begin{table}[t]
\centering
\caption{Statistics of the curated instrument transfer set.}
\label{tab:dataset_stats}
\small
\resizebox{1.0\linewidth}{!}{
\begin{tabular}{llcc}
\toprule
\textbf{Cluster} & \textbf{\# Instruments} & \textbf{Samples} & \textbf{Duration (min)} \\
\midrule
$C_{\text{High}}$ & 5: \textit{Vn, Fl, Ob, Cl, Tp} & 597 & 119.4 \\
$C_{\text{Mid}}$ & 4: \textit{Va, Bn, Sax, Hn} & 391 & 78.2 \\
$C_{\text{Low}}$ & 4: \textit{Db, Vc, Tu, Tb} & 333 & 66.6 \\
\midrule
\textbf{Total} & \textbf{13} & \textbf{1,321} & \textbf{264.2} \\
\bottomrule
\end{tabular}
}
\vspace{-5mm}
\end{table}

%% file: sec/3_experiment.tex
\section{Experiment}

\subsection{Experimental setup}

\noindent\textbf{Datasets.}
We consolidated samples from the URMP~\cite{Li18-URMP} and Solos~\cite{Montesinos20-Solos} datasets, covering 13 instrument types. All recordings were resampled to 44.1\,kHz and segmented into 12-second clips. To eliminate non-musical segments, we performed automated filtering via CLAP score~\cite{Wu23-CLAP}, excluding samples with a cosine similarity $>0.5$ to non-musical textual queries (e.g., ``applaud'', ``talk'').
Text prompts for training were constructed using the template: \textit{``a monophonic single melody of \textless instrument\textgreater\ playing solo, \textless instrument family\textgreater\ instrument.''}
We applied lexical augmentation to the descriptive modifiers
by uniformly sampling 
one from two synonymous alternatives (e.g., \textit{monophonic} and \textit{single}).
The training split (32.8 hrs) comprises two subsets: (1) an \textit{instrument reconstruction set} to optimize SAO-ControlNet, and (2) an \textit{instrument transfer set}
synthesized in Section \ref{sec:data_construction} to train \ours.
The evaluation set comprises 2,400 text--audio pairs, formed by pairing 100 held-out samples per instrument with all other candidate instruments within their respective predefined pitch clusters.

\noindent\textbf{Metrics.}
We employed the following objective metrics:
(1) CLAP score~\cite{Wu23-CLAP} to quantify timbral fidelity via text--audio alignment;
(2) Chroma score to measure structural consistency using chromagram cosine similarity, with a 25-cent resolution to capture fine-grained expressive details;
(3) $F1_{\text{MIDI}}$ score to evaluate score-level agreement between MIDI transcriptions extracted from both source and generated audio using YourMT3+~\cite{Chang24-YourMT3};
and (4) Kernel Audio Distance (KAD)~\cite{Chung2025KAD} to assess overall audio quality based on MERT~\cite{Li2024MERT} embeddings.
For subjective evaluation, 22 participants rated 20 test items on a 5-point Likert scale across  four dimensions:
timbral fidelity (TIM), timbral naturalness (NAT), structural fidelity (STR), and overall quality (QUL).

\noindent\textbf{Implementation details.}
Built upon SAO, our system adopts a 12-second generation length ($256 \times 64$ latents) to address computational constraints.
Training followed a two-stage procedure using the AdamW optimizer and Mean Squared Error loss: (1) SAO-ControlNet was trained for 1,200 epochs (batch size 384); and (2) \ours was trained for 400 epochs (batch size 64) with SAO-ControlNet frozen.
The first stage decayed the learning rate from $10^{-4}$ to $10^{-5}$ after a 5-epoch warmup, while the rate remained fixed at $10^{-5}$ during the second stage.

\input{tab/table_ablation_bins}
\input{tab/table_ablation}
\input{tab/table_ttm_editing}

\subsection{Results and analysis}

\noindent\textbf{Effect of control resolution.}
As a preliminary study,
we examine the performance trade-offs of
structural control resolution in SAO-ControlNet during timbre transfer.
Table~\ref{tab:ablation_bins} reveals divergent trends in CLAP and Chroma scores relative to control granularity: while higher bin counts improve overall structural consistency, they simultaneously degrade timbral fidelity.
As a pragmatic compromise, we adopted the 144-bin $f_0$ and 32-bin RMS configuration for subsequent experiments, which serves as the optimal operating point between these competing metrics.

\begin{figure}[t!]
  \centering
  \includegraphics[width=\linewidth]{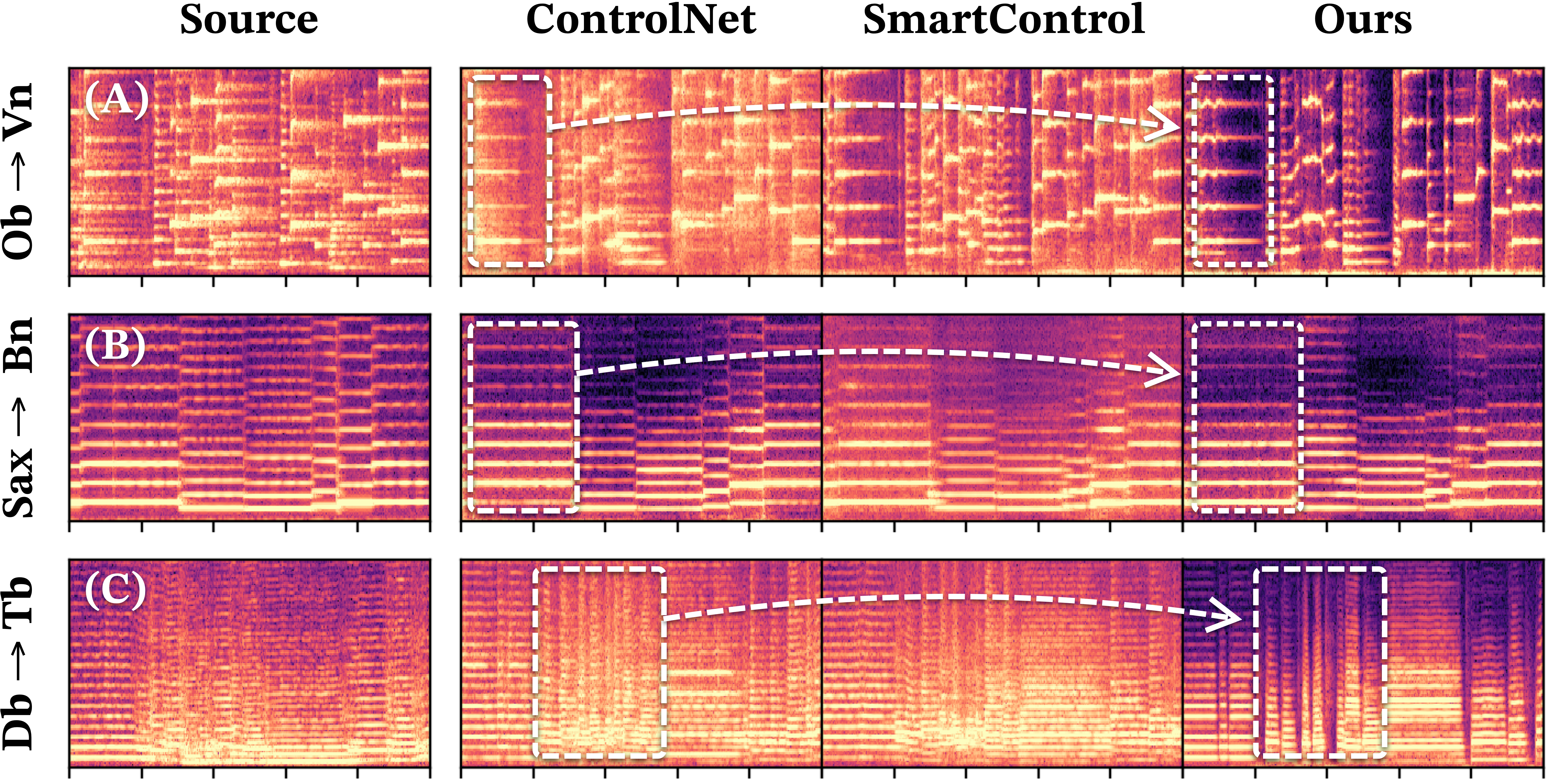}
  \caption{Visual comparison of timbre transfer. \ours achieves high-fidelity target timbres by (A) synthesizing missing violin-specific vibrato, (B) mitigating high-frequency vibrato artifacts, and (C) producing clear trombone-specific attack transients.}
  \label{fig:demo_examples}
  \vspace{-10pt}
\end{figure}

\noindent\textbf{Comparison with ControlNet-based baselines.}
Table~\ref{tab:main_results} demonstrates the efficacy of \ours, highlighting its capability to enhance timbral fidelity with minimal compromise to score-level content preservation
within the ControlNet framework.
In terms of timbral quality, our method achieves the highest CLAP score of 0.490, matching the objective upper bound of the SAO backbone, along with the highest subjective timbral fidelity (TIM) of 3.582 and timbral naturalness (NAT) of 3.484, surpassing both ControlNet (without CSPs and TG-CSPs) and SmartControl (without TG-CSPs).
This confirms that our approach successfully tailors source expressive details to 
the target instrument's identity; the adaptive refinement illustrated in Figure~\ref{fig:demo_examples} yields 
perceptually authentic audio.
Regarding score-level content preservation, although our method shows a marginal reduction in $F1_{\text{MIDI}}$ score to 0.302 compared to ControlNet, it notably achieves the highest subjective structural fidelity (STR) of 4.148.
We attribute this objective--subjective discrepancy to
the adaptive refinement of instrument-specific expressive nuances such as attack transients and pitch glides. While these subtle deviations may marginally impact transcription results, \ours prioritizes perceptual naturalness over rigid structural matching, at the cost of a slight decline in $F1_{\text{MIDI}}$.
For overall audio quality, our method yields the lowest KAD of 0.495 among ControlNet-based baselines and achieves the highest overall quality (QUL) score of 3.307, demonstrating its ability to bridge the distributional gap relative to real recordings.
Particularly in terms of objective performance, this stands in contrast to SmartControl, which exhibits a slight performance degradation in KAD compared to ControlNet.
Such limited efficacy confirms that simply scaling the ControlNet layer-wise outputs is insufficient for modeling the relative influence of heterogeneous controls, due to their deep entanglement within its hidden space.

\noindent\textbf{Comparison with SoTA inference-time editing models.}
Table~\ref{tab:ttm_editing} highlights the advantages of our ControlNet-based finetuning over inference-time editing paradigms, particularly in timbre transfer where score-level content preservation is paramount.
Two inference-time editing baselines, MusicMagus~\cite{Zhang24-MusicMagus} and ZETA~\cite{Manor24-ZETA},
yield suboptimal performance in both CLAP and $F1_{\text{MIDI}}$ scores, with a particularly pronounced gap in the latter.
Even after finetuning on our \textit{instrument reconstruction set} ($\dagger$), they fail to reach \ours across all objective metrics. This indicates that inference-time editing paradigms,
which forgo a training phase on explicit conditioning,
are highly sensitive to hyperparameters such as the diffusion inversion timestep, which governs the noise level applied to source latents before target-conditioned denoising.
This inherent sensitivity makes it difficult to strike an optimal balance between timbral fidelity and structural consistency, often leading to either timbral distortion or score-level content collapse.
Crucially, this sensitivity further manifests in overall audio quality:
\ours yields a KAD of 0.495,
whereas MusicMagus$^{\dagger}$, the best-performing baseline in this metric,
remains at a significantly higher KAD of 1.408,
demonstrating a substantial improvement in bridging the audio quality gap
toward real-world recordings.
This confirms that without a target-adaptive mechanism, high hyperparameter sensitivity inevitably compounds into such audio quality degradation.
Ultimately, these findings demonstrate that the structural stability conferred by ControlNet-based finetuning provides a more resilient foundation
for timbre transfer than inference-time editing.

%% file: tab/table_ablation_bins.tex
\begin{table}[t!]
    \centering
    \caption{Trade-off of SAO-ControlNet: Finer control signals improve structural consistency but degrade timbral accuracy.}
    \label{tab:ablation_bins}
    \vspace{-0.2cm}
    \small
    \renewcommand{\arraystretch}{1.3}
    \setlength{\aboverulesep}{2pt}
    \setlength{\belowrulesep}{2pt}

    \tikzset{
        floating_dash/.style={
            draw, dashed, overlay,
            inner sep=0pt,
            minimum width=3.4em,
            minimum height=1.35em,
            line width=0.8pt, color=black!80
        }
    }

    \resizebox{\columnwidth}{!}{%
        \begin{tabular}{c | ccc | ccc}
            \toprule
            \diagbox[width=6.0em, height=2.0em]{\raisebox{-2pt}{\hspace*{2pt}\textbf{$f_0$}}}{\raisebox{2pt}{RMS}\hspace*{2pt}} &
            16 & 32 & 64 & 16 & 32 & 64 \\

            \midrule

            288 &
            \cB{50}0.455 & \cB{40}0.452 & \cB{20}0.447 &
            \cR{70}0.831 & \cR{90}\textbf{0.841} & \cR{80}\underline{0.839} \\

            144 &
            \cB{60}0.457 &
            \begin{tikzpicture}[baseline=(char.base)]
                \node[inner sep=0pt, text depth=0pt] (char){\cB{80}\underline{0.463}};
                \node[floating_dash] at (char.center) {};
            \end{tikzpicture} &
            \cB{20}0.447 &
            \cR{45}0.808 &
            \begin{tikzpicture}[baseline=(char.base)]
                \node[inner sep=0pt] (char) {\cR{45}0.809};
                \node[floating_dash] at (char.center) {};
            \end{tikzpicture} &
            \cR{50}0.811 \\

            72 &
            \cB{65}0.458 & \cB{90}\textbf{0.466} & \cB{45}0.453 &
            \cR{5}0.767 & \cR{5}0.767 & \cR{10}0.770 \\

            \bottomrule
        \end{tabular}%
    }
    \par\vspace{4pt}
    {\raggedright \scriptsize
    \text{ \textcolor{softBlue}{\rule[-0.1em]{0.8em}{0.8em}} CLAP $\uparrow$ \quad \textcolor{softRed}{\rule[-0.1em]{0.8em}{0.8em}} \text{Chroma} $\uparrow$} \par}

    \vspace{-4mm}
\end{table}

%% file: tab/table_ablation.tex
\begin{table*}[t]
  \caption{Evaluation of timbre transfer performance. Our framework achieves the best performance of timbral accuracy and naturalness while preserving score-level content comparable to ControlNet. Bold and underlined scores denote best and second-best.}
  \label{tab:main_results}
  \centering
  \footnotesize
  \setlength{\tabcolsep}{4pt}

  \begin{tabularx}{\textwidth}{l >{\centering\arraybackslash}p{2cm} | YYY | YYYY }
    \toprule
    \multirow{2}{*}{\textbf{Model}} & \multirow{2}{*}{\textbf{Condition}} & \multicolumn{3}{c|}{\textbf{Objective Metrics}} & \multicolumn{4}{c}{\textbf{Subjective Metrics}} \\
    \cmidrule{3-5} \cmidrule{6-9}
    & & \textbf{CLAP} $\uparrow$ & \textbf{$F1_{\text{MIDI}}$} $\uparrow$ & \textbf{KAD} $\downarrow$ & \textbf{TIM} $\uparrow$ & \textbf{NAT} $\uparrow$ & \textbf{STR} $\uparrow$ & \textbf{QUL} $\uparrow$ \\
    \midrule
    \textcolor{gray}{\textit{SAO}} & \textcolor{gray}{\textit{Text}} & \textcolor{gray}{\textit{0.490}} & \textcolor{gray}{\textit{0.004}} & \textcolor{gray}{\textit{0.331}} & \textcolor{gray}{\textit{3.452}} & \textcolor{gray}{\textit{3.259}} & \textcolor{gray}{\textit{1.439}} & \textcolor{gray}{\textit{3.034}} \\
    \midrule
    ControlNet & \textit{Text, $f_0$, RMS} & 0.463 & \textbf{0.309} & \underline{0.512} & 3.164 & 3.136 & \underline{3.998} & 2.875 \\
    SmartControl & \textit{Text, $f_0$, RMS} & \underline{0.471} & 0.293 & 0.520 & \underline{3.418} & \underline{3.366} & 3.991 & \underline{3.107} \\
    \textbf{\ours} (ours) & \textit{Text, $f_0$, RMS} & \textbf{0.490} & \underline{0.302} & \textbf{0.495} & \textbf{3.582} & \textbf{3.484} & \textbf{4.148} & \textbf{3.307} \\
    \bottomrule
  \end{tabularx}
  \vspace{-2mm}
\end{table*}

%% file: tab/table_ttm_editing.tex
\begin{table}[t]
  \caption{Comparison results with unsupervised inference-time editing baselines ($^{\dagger}$ finetuned on SAO-ControlNet training set). Bold and underlined scores denote best and second-best.}
  \label{tab:ttm_editing}
  \centering
  \footnotesize
  \setlength{\tabcolsep}{0pt}
  \begin{tabular*}{\columnwidth}{@{\extracolsep{\fill}} l c c c }
    \toprule
    \textbf{Model} &
    \textbf{CLAP $\uparrow$} &
    \textbf{$F1_{\text{MIDI}} \uparrow$} &
    \textbf{KAD $\downarrow$} \\
    \midrule
    MusicMagus & 0.357 & 0.157 & 1.577 \\
    MusicMagus$^{\dagger}$ & 0.387 & 0.176 & \underline{1.408} \\
    ZETA (w/ SAO) & 0.472 & 0.167 & 1.943 \\
    ZETA$^{\dagger}$ (w/ SAO) & \underline{0.486} & \underline{0.181} & 1.897 \\
    \midrule
    \textbf{\ours} (ours) & \textbf{0.490} & \textbf{0.302} & \textbf{0.495} \\
    \bottomrule
  \end{tabular*}
  \vspace{-4mm}
\end{table}

%% file: sec/4_conclusion.tex
\section{Conclusion}

We propose \ours to achieve high timbral fidelity across diverse transfer scenarios. By coupling the ControlNet framework with a semi-automatic data construction pipeline, our method modulates heterogeneous structural controls via text prompts to match target timbre, demonstrating superior timbre transfer capabilities during evaluation.
Despite these advantages, \ours remains limited to monophonic audio and does not preserve original spatial cues (e.g., reverberation). 
Future work will extend adaptive control to polyphonic generation and integrate disentangled representations to preserve spatial characteristics.